\documentclass[10pt]{article}
\usepackage[utf8]{inputenc}
\usepackage{epsfig}
\usepackage{amsmath}
\usepackage{graphicx,psfrag}
\usepackage{graphicx}
\usepackage{float}
\usepackage{subfigure}

\begin{document}

\title{
\it Soliton propagation in lossy optical fibers \\ \vspace{0.5cm}
\bf Propagação de sólitons em fibras óticas dissipativas}

\author{Caroline Dall'Agnol$^1$; Paulo Laerte Natti$^2$; Eliandro Rodrigues Cirilo$^3$; \\
       Neyva Maria Lopes Romeiro$^4$; \'Erica Regina Takano Natti$^5$ \\ \\	
 {\small $^1$Profa. Ms. UTFPR, Câmpus Dois Vizinhos, Pr, Brasil; E-mail: carol{\_}dall@hotmail.com}\\
 {\small $^2$Corresponding author. Prof. Dr. Mathematics Depto, UEL, Pr, Brasil;  E-mail: plnatti@uel.br}\\
 {\small Tel: 55-43-33714226. Fax: 55-43-33714216}\\
 {\small $^3$Prof. Dr. Mathematics Depto, UEL, Pr, Brasil;  E-mail: ercilo@uel.br}\\
 {\small $^4$Profa. Dra. Mathematics Depto, UEL, Pr, Brasil; E-mail: nromeiro@uel.br}\\
 {\small $^5$Profa. Dra. PUC, Câmpus Londrina, Pr, Brasil; E-mail: erica.natti@pucpr.br}}

\date{}

\maketitle

{\centerline{\bf{\large Abstract}}}

\noindent
In this work we study the propagation of solitons in lossy optical fibers. The main objective of this work is to study the loss of energy of the soliton wave during propagation and  then to evaluate the impact of this loss on the transmission of the soliton signal. In this context, a numerical scheme was developed to solve a system of complex partial differential equations (CPDE) that describes the propagation of solitons in optical fibers with loss and nonlinear amplification mechanisms. The numerical procedure is based on the mathematical theory of Taylor series of complex functions. We adapted the Finite Difference Method (FDM) to approximate derivatives of complex functions. Then, we solve the algebraic system resulting from the discretization, implicitly, through the relaxation Gauss-Seidel method (RGSM). The numerical study of CPDE system with linear and cubic attenuation showed that soliton waves undergo attenuation, dispersion, and oscillation effects. On the other hand, we find that by considering the nonlinear term (cubic term) as an optical amplification, it is possible to partially compensate the attenuation of the optical signal. Finally, we show that a gain of 9$\%$ triples the propagation distance of the fundamental soliton wave, when the dissipation rate is 1$\%$.

\vspace{0.5cm}
{\centerline{\bf{\large Resumo}}}

\noindent
Neste trabalho estudamos a propagação de solitons em fibras ópticas com perdas. O principal objetivo deste trabalho é estudar a perda de energia da onda soliton durante a propagação e avaliar o impacto dessa perda na transmissão do sinal soliton. Neste contexto, um esquema numérico foi desenvolvido para resolver um sistema de equações diferenciais parciais complexas (EDPC), que descreve a propagação de solitons em fibras óticas com mecanismos de perdas e de amplificações não-lineares. O procedimento numérico é baseado na teoria matemática das séries de Taylor para funções complexas. Adaptamos o método de diferenças finitas (MDF) para aproximar derivadas de funções complexas. Em seguida, resolvemos o sistema algébrico resultante da discretização, implicitamente, por meio do método de Gauss-Seidel com relaxamento (MGSR). O estudo numérico do sistema de EDPC com atenuação linear e cúbica mostrou que ondas soliton sofrem efeitos de atenuação, dispersão e oscilação. Por outro lado, verificamos que ao considerar o termo não linear (termo cúbico) como uma amplificação ótica é possível compensar parcialmente a atenuação do sinal ótico. Finalmente, mostramos que um ganho de 9 $\%$ triplica a distância de propagação da onda soliton fundamental, quando a taxa de dissipação é de 1 $\%$.\\
\noindent{\bf Palavras-chave}: Comunicação Óptica. Soliton. Diferenças Finitas. Dissipação. Amplificação Não Linear.

\twocolumn
\section*{Introduction}

One of the essential activities of the human condition is communication. The exchange of information has become very important for both personal and professional life. This exchange of information can occur analogically or digitally through various means such as coaxial cables, microwave, radio frequencies, infrared and optical fibers. With the Internet access facilitated through cell phones, notebooks and tablets, we see the daily growth of information traffic, which forces communication systems to have greater transmission capacity, high bandwidth, efficiency and speed with no significant energy losses and with low costs.

Among the means of data transmission cited, fiber optic communication systems are the most suitable for these requirements, since they have low attenuation factor, high bandwidth and low production costs. In addition, due to its lightweight, flexible and low volume format, optical fibers have low storage costs. Another important feature of optical fibers is that they have high electrical resistance, which makes them a highly insulating medium, preventing external electromagnetic interference and guaranteeing better transmission quality, with secrecy, among other advantages.

On the other hand, the physical characteristics of optical fibers are very important for preserving the quality of the transmitted signal, so that inomogeneities, diffusion of hydrogen molecules, bubbles, variations in diameter and roughness perturb the propagation of signals, generating noise and loss power. Therefore, in the last decades several experiments have been carried out to compensate the effects of dispersion and nonlinearities in optical communication systems over long distances in order to increase their data transmission capacity. One of the most important innovations in the field of communication technology, able to overcome these difficulties, is based on the concept of optical solitons.

Solitons are optical pulses capable of keeping their shape unaltered in non-linear and dispersive media, such as in optical fibers. The principle of propagation of the solitons in optical fibers is based on the perfect balance between the Group Velocity Dispersion (GVD) and the Kerr effect due to the non-linearity of the medium (TAYLOR, 1992; MENYUK; SCHIEK; TORNER, 1994; AGRAWAL, 2019).

Historically, Smith et al. (1996) showed that solitons could propagate in fibers with periodic variation of the GVD, even if the mean dispersion was practically null, so a new idea emerged for systems with solitons: the systems with managed dispersion. Other experiments were performed by Fukuchi et al. (2001) on mono-channel and WDM systems, aiming to reach transmission capacity above 1 Tb/s. Later, Algety Telecom, based in Lannion, France, made the practical use of solitons a reality when developing submarine telecommunications equipment based on the transmission of optical solitons (ALGETY, 2002).

Currently, much research has been carried out to find a soliton transmission system capable of competing commercially with current communication systems, that is, offering high transmission rates at low cost (CHEMNITZ et al., 2017; LUO et al., 2017; ZAJNULINA et al., 2017; EFTEKHAR et al., 2019; WANG et al., 2019). 

In this context, this work performs numerical studies describing the propagation of solitons in dielectric optical fibers, with emphasis on the study of power loss and the evaluation of the impact of this dissipation in the transmission of the soliton signal.

\section*{Solitons in optical guides}

This section studies the coupled non-linear system of complex partial differential equations (CPDE), obtained from Maxwell’s equations, which describe the longitudinal propagation of two coupled electromagnetic waves (fundamental and second harmonic modes) in ideal $\chi^{(2)}$ dielectric optical fibers (GALLÉAS; YMAI; NATTI; NATTI, 2003). This CPDE system is given by
	{\small{\begin{eqnarray} 
	\label{maxwell} 
	i\frac{\partial a_1}{\partial \xi}\ - \frac{r}{2}\frac{\partial^2 a_1}{\partial s^2} + 
	a_1 ^{\ast} a_2 \;{\exp}(-i \beta \xi)=0 \nonumber \\ \\ 
	i\frac{\partial a_2}{\partial \xi}\ - i \delta \frac{\partial a_2}{\partial s} - 
	\frac{\alpha}{2}\frac{\partial^2 a_2}{\partial s^2}+ a_1 ^2 \;{\exp}(i \beta \xi)=0 , \nonumber
	\end{eqnarray}}}
	
\noindent
where $i=\sqrt{-1}$ is the imaginary unit, $a_1(\xi,s)$ and $a_2(\xi,s)$ are complex variables that represent the normalized amplitudes of the electrical fields of the fundamental and second harmonic waves, respectively, with 
$a_1^{*}(\xi,s)$ as its complex conjugates. The independent variable $s$ has spatial character, and the independent variable $\xi$ has temporal character. The real parameters $\alpha$, $\beta$, $\delta$ and $r$ are related with the dielectric properties of the optical fiber 
(YAMAI; GALLÉAS; NATTI; NATTI, 2004; QUEIROZ; NATTI; ROMEIRO; NATTI, 2006; CIRILO; NATTI; ROMEIRO; TAKANO NATTI, 2008).

The CPDE system (\ref{maxwell}) presents ideal soliton solutions (GALLÉAS; YMAI; NATTI; NATTI, 2003) given by
	{\scriptsize{\begin{eqnarray} \label{a1}
a_1(\xi,s) = \pm\frac{3}{2(\alpha-2r)}\sqrt{\alpha r}\left(\frac{\delta^2}{2\alpha-r}+\beta\right) \\ 
  sech^2\left[\pm\sqrt{\frac{1}{2(2r-\alpha)}\left(\frac{\delta^2}{2\alpha-r}+\beta\right)}\left(s-\frac{r\delta}{2\alpha-r}\xi\right)\right] \nonumber \\
 {\exp}\left\{i\left[\frac{r\delta^2(4r-5\alpha)}{2(2\alpha-r)^2(2r-\alpha)}-\frac{r\beta}{2r-\alpha}\right]\xi-\frac{i\delta}{2\alpha-r}s\right\} \nonumber 
	\end{eqnarray}}}
	{\scriptsize{\begin{eqnarray} \label{a2}
a_2(\xi,s) = \pm\frac{3r}{2(\alpha-2r)}\left(\frac{\delta^2}{2\alpha-r}+\beta\right) \\ 
  sech^2\left[\pm\sqrt{\frac{1}{2(2r-\alpha)}\left(\frac{\delta^2}{2\alpha-r}+\beta\right)}\left(s-\frac{r\delta}{2\alpha-r}\xi\right)\right] \nonumber \\
 {\exp}\left\{i\left[\frac{2r\delta^2(4r-5\alpha)}{2(2\alpha-r)^2(2r-\alpha)}-\frac{2r\beta}{2r-\alpha}+\beta\right]\xi-\frac{2i\delta}{2\alpha-r}s\right\}. \nonumber
	\end{eqnarray}}}

\noindent
The existence of these solutions (YAMAI; GALLÉ\-AS; NATTI; NATTI, 2004) is conditioned by the fact that the hyperbolic secant function argument is real and by the non-existence of singularities in (\ref{a1}) and (\ref{a2}).

As we know, propagation of soliton-type waves in real optical fibers does not occur in the same way as in ideal optical fibers (CIRILO; NATTI; ROMEIRO; NATTI; OLIVEIRA, 2010). There are many processes that can cause disturbances in the propagation of real solitons: connection/fusion between optical fibers (WANG; ZHOU; XU; YANG; ZHANG, 2019; OLIVEIRA; NATTI; CIRILO; RO\-MEIRO; NATTI, 2020), Rayleigh scattering (PAL\-MIERI; SCHENATO, 2013), high-order dispersion and high-order nonlinearities (TRIKI; BISWAS; MI\-LOVIC; BELIC, 2016), soliton self-steepening, Raman effect and self-frequency shift (WEN; YANGBAO; SHI; FU, 2018), polarization-mode dispersion (KUMAR; RAO, 2012), nonlinear phase noise (YUSHKO; REDYUK; FEDORUK; TURITSYN, 2014), among other processes (ASHRAF; AHMAD; YOUNIS; ALI; RIZVI, 2017). 

It should be observed that the perturbed coupled nonlinear Schrödinger differential equations systems, which describe wave propagation in real optical media, do not present analytical solution. In the literature there are several numerical approaches whose objective is to describe the propagation of perturbed solitons in dielectric environments, most of them using the finite difference method (ISMAIL; ASHI, 2016; OLIVEIRA; NATTI; CIRILO; ROMEI\-RO; NATTI, 2020) or the finite element method (ISMAIL, 2008; KARCZEWSKA; ROZMEJ; SZCZE\-CINSKI; BOGONIEWICZ, 2016).

In this context, in order to observe the loss of energy caused by these processes in the soliton wave propagation, we add two perturbative terms to equation (\ref{maxwell}), ie,
{\small{\begin{eqnarray} \label{perturbada} 
	i\frac{\partial a_1}{\partial \xi}\ - \frac{r}{2}\frac{\partial^2 a_1}{\partial s^2} 
	+ a_1 ^{\ast} a_2 \;{\exp}(-i \beta \xi) \nonumber \\
	= i\mu\;a_1 + i\kappa\;|a_1|^{2}\; a_1 \nonumber \\
\end{eqnarray}
\vspace{-0.8cm}
\begin{eqnarray} 
	i\frac{\partial a_2}{\partial \xi}\ - i \delta \frac{\partial a_2}{\partial s} 
	- \frac{\alpha}{2}\frac{\partial^2 a_2}{\partial s^2}+ a_1 ^2 \;{\exp}(i \beta \xi) \nonumber \\
	=i\mu\;a_2 + i\kappa\;|a_2|^{2}\; a_2 \hspace{0.25cm}, \nonumber
\end{eqnarray}}}

\noindent
where $i\mu\;a_1$ and $i\mu\;a_2$ are linear dissipative terms, while $i\kappa\;|a_1|^{2}\; a_1$ and $i\kappa\;|a_2|^{2}\; a_2$ are nonlinear dissipative or amplification terms. In this case the propagating wave is not a soliton, and only numerical solutions are possible. 
 
In (CIRILO; NATTI; ROMEIRO; NATTI; OLI\-VEIRA, 2010) a numerical procedure was developed to solve the propagation of soliton waves in ideal optical fibers described by the PDE system (\ref{maxwell}). The procedure is based on the finite difference method for complex functions and relaxation Gauss-Seidel method. In this work we perform adaptations in the numerical procedure developed, in order to solve the system (\ref{perturbada}). In this context, we made approximations in order to obtain an Implicit Method, because the resulting linear system (in complex variables) became well-conditioned. We chose to solve the resulting linear system by the Relaxation Gauss-Seidel method, which accelerates the convergence (CIRILO; NATTI; ROMEIRO; NATTI, 2008). Note that the resolution of the complex linear system can be performed by other procedures, such as Cholesky decomposition, conjugate gradiente, tridiagonal matrix algorithm (TDMA), modified strongly implicit procedure (MSI), among others. We chose the Relaxation Gauss-Seidel method because of its mathematical simplicity and easy computational implementation (SMITH, 2004; CIRILO; NATTI; RO\-MEIRO; NATTI; OLIVEIRA, 2010; ROMEIRO; CASTRO; CIRILO; NATTI, 2011; PARDO; NATTI; ROMEIRO; CIRILO, 2012; SAITA et al., 2017; ROMEIRO; MANGILI; COSTANZI; CIRILO; NA\-TTI, 2017; CIRILO; BARBA; NATTI; ROMEIRO, 2018; CIRILO; PETROVSKI; ROMEIRO; NATTI, 2019; OLIVEIRA; NATTI; CIRILO; ROMEIRO; NATTI, 2020). This numerical procedure is presented in the next section.

\section*{Numerical model}

The system (\ref{perturbada}) is numerically resolved in domain $\xi\times s=[0,T] \times [-L,L]$, where $T,L\in {R}$. By discretizing the variables $a_1(\xi,s)\equiv a_1(k+1,j)$ and $a_2(\xi,s)\equiv a_2(k+1,j)$ for $k=0,1,...,k_{max}$ and $j=1,2,...,ni$, where $k_{max}$ is denominated the last advance in $\xi$ and $ni$ the maximum number of points in $s$, the propagation domain of the soliton waves is defined by a discretized computational grid of $k_{max}\times ni$ points, as represented in figure 1.
 
\begin{figure}[!ht]
\begin{center}
\caption{Computational domain of propagation of soliton waves.}
\includegraphics[scale=0.6]{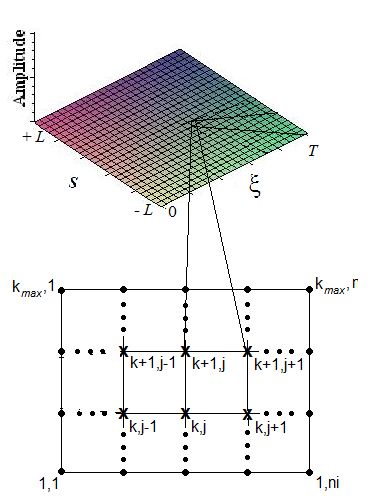}
	\label{dominio}
\center{{\bf{Source:}} The Authors}
\end{center}
\end{figure}

Thus, by means of the finite difference method, by approximating the temporal derivatives by progressive differences, and the spatial derivatives by central differences, we can rewrite (\ref{perturbada}) as:
  {\small{\begin{eqnarray} \label{a1ddp}
a_{1}(k+1,j)=\left(\frac{1}{^{1} A_{p}}\right)[\:^{1}A_{w}\;a_{1}(k+1,j-1) \nonumber \\
+^{1}A_{e}\;a_{1}(k+1,j+1)-\;a_1^{*}(k,j)\;a_2(k,j)\;{\exp}(-i\beta\xi) \nonumber \\
+^{1}A_{p_{0}}\;a_{1}(k,j)]
	\end{eqnarray}}}	
	{\small{\begin{eqnarray} \label{a2ddp}
a_{2}(k+1,j)=\left(\frac{1}{^{2}A_{p}}\right)[\:^{2}A_{w}\;a_{2}(k+1,j-1) \nonumber \\
+^{2}A_{e}\;a_{2}(k+1,j+1)-\;a_1^2(k+1,j)\;{\exp}(i\beta\xi) \nonumber \\ 
+\;^{2}A_{p_{0}}\;a_{2}(k,j)],
	\end{eqnarray}}}
	
\noindent where
{\scriptsize{$$\begin{array}{lllllll} 
^{1}A_{p}= i\left(\frac{1}{\Delta\xi}-\mu - \kappa|a_{1_{k+1,j}}|^{2} \right) + \displaystyle\frac{r}{(\Delta s)^2} 
& & ^{1}A_{w}=\:^{1}A_{e} \\ \\
^{2}A_{e}=\displaystyle\frac{\alpha}{2(\Delta s)^2}+\displaystyle\frac{i\delta}{2\Delta s}  
& & ^{1}A_{w}=\displaystyle\frac{r}{2(\Delta s)^2} \\ \\
^{2}A_{p}= i\left(\frac{1}{\Delta\xi} - \mu - \kappa|a_{2_{k+1,j}}|^{2} \right) + \displaystyle\frac{\alpha}{(\Delta s)^2} 
& & ^{1}A_{p_{0}}=\:^{2}A_{p_{0}} \\ \\
^{2}\hspace{-0.05cm}A_{w}=\displaystyle\frac{\alpha}{2(\Delta s)^2}-\displaystyle\frac{i\delta}{2\Delta s} 
& & ^{1}\hspace{-0.013cm}A_{p_{0}}=\displaystyle \frac{i}{\Delta\xi}. \\ \\
\end{array}$$}}

Applying the relaxation Gauss-Seidel Method (CIRILO; NATTI; ROMEIRO; NATTI, 2008), we can compute iteratively $a_{1}(k+1,j)^{n+1}$ through the equations given below
{\small{\begin{eqnarray} 
a_{1}(k+1,j)^{(n+1)}=\frac{+^{1}A_{w}\;a_{1}(k+1,j-1)^{(n+1)}}{^{1}A_{p}} \nonumber \\
+\frac{\;^{1}B_{j}+\;^{1}A_{e}\;a_{1}(k+1,j+1)^{(n)}}{^{1}A_{p}},
\end{eqnarray}}}

\noindent until the stop criterion is satisfied, ie,
{\small{\begin{eqnarray}
\max_{2 \leq j \leq n_{i-1}}|a_{1}(k+1,j)^{(n+1)}-a_{1}(k+1,j)^{(n)}|<10^{-6}
\end{eqnarray}}}

\noindent where 
{\small{\begin{eqnarray}
^{1}B_{j}&=&\:^{1}A_{p_{0}}\;a_{1}(k,j)^{(n+1)} \nonumber \\
&-&a_1^{\ast}(k,j)^{(n+1)}\;a_2 (k,j)^{(n+1)}\;{\exp}(-i\beta\xi). \nonumber
\end{eqnarray}}}

\noindent
Similarly $a_{2}(k+1,j)^{(n+1)}$ is calculated.

\subsection*{Numerical simulations for solitons with dissipation}

In this section, in all the simulations of system (4), we assume for the dielectric parameters the following values: $r=-1$, $\beta=-0,5$, $\alpha=-0,25$ and $\delta=-0,1$. These values are compatible with those measured in real optical fibers (ARTIGAS, 1999). On discretization, we define $\Delta s=1,0\times10^{-1}$ and $\Delta\xi=1,0\times10^{-3}$ for the variables $s$ and 
$\xi$, respectively. The intervals of variation of $s$ and $\xi$ were adapted to each case, so that the plots of the propagated solitons remained in the computational domain, as represented in the figure 1.

Figure 2 shows the propagation of the modes $|a_1(\xi,s)|$ and $|a_2(\xi,s)|$ of soliton wave packet under ideal conditions, without dissipative or gain terms, ie, with $\mu=\kappa=0$.

\begin{figure}
\begin{center}
\caption{Propagation of modes $|a_1(\xi,s)|$ and $|a_2(\xi,s)|$ of ideal soliton wave ($\mu=\kappa=0$).}
\subfigure[Mode $|a_1(\xi,s)|$]{\includegraphics[scale=0.37]{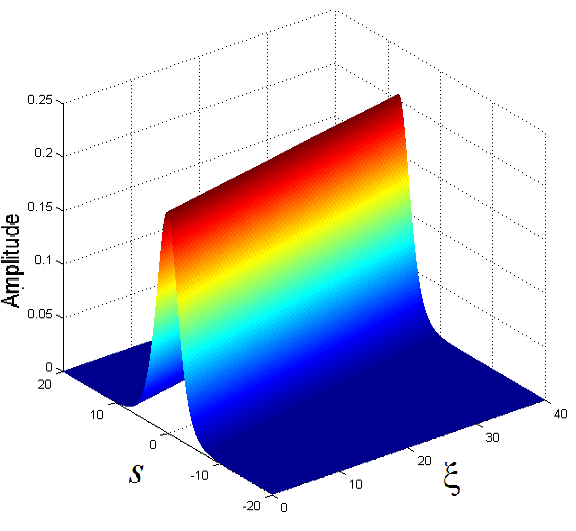}
  \label{}}\qquad \qquad
\subfigure[Mode $|a_2(\xi,s)|$]{\includegraphics[scale=0.37]{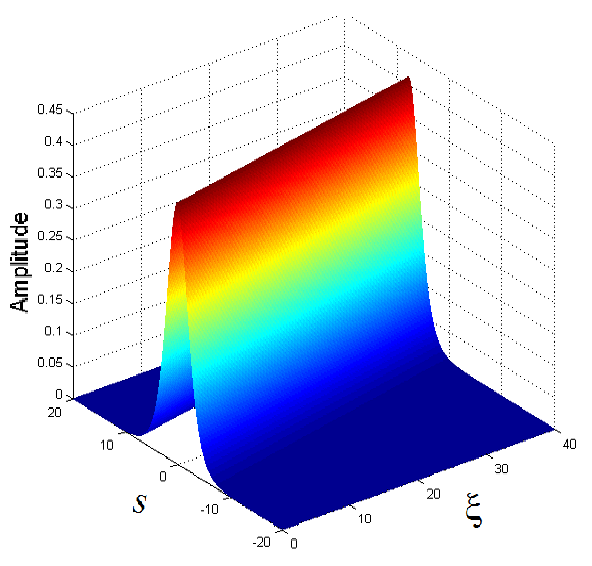}
   \label{}} 
\label{ideal}  
\center{{\bf{Source:}} The Authors}
\end{center}
\end{figure}

\begin{figure}
\begin{center}
\caption{Profile of $|a_1(\xi,s)|$ and $|a_2(\xi,s)|$ modes of the ideal soliton wave ($\mu=\kappa=0$).}
\includegraphics[scale=0.25]{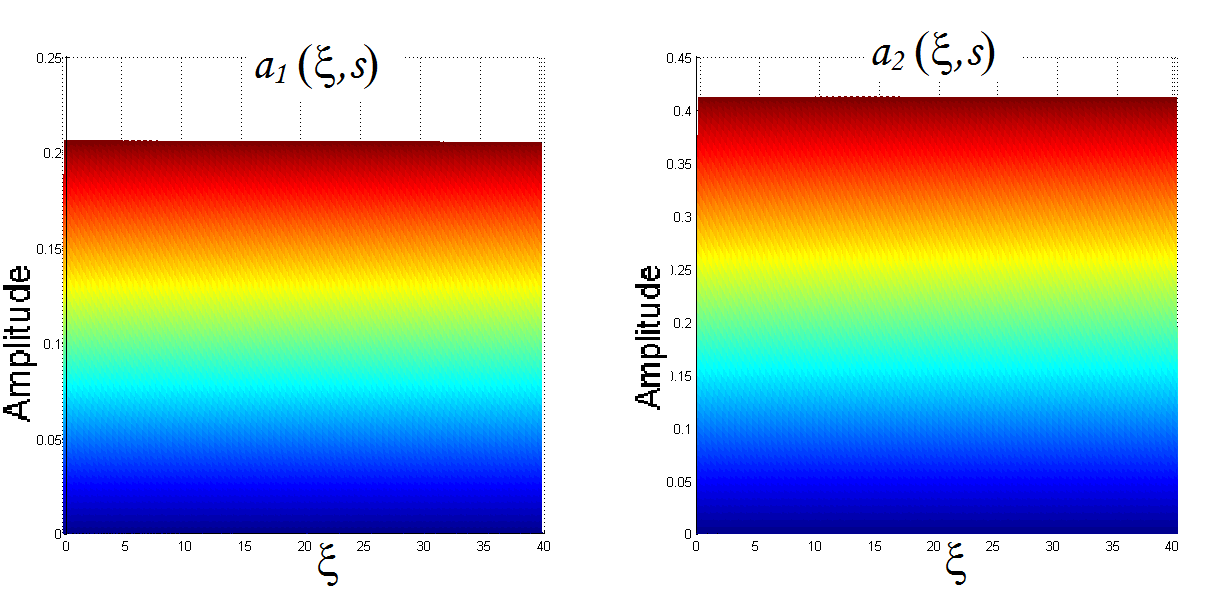}
\label{idealp}  
\center{{\bf{Source:}} The Authors}
\end{center}
\end{figure}

Note that propagation occurs without variations in wave amplitudes, which can best be observed in figure 3, where the evolution of the amplitudes as a function of the time variable $\xi$ is visualized.

In the following simulation we add the linear dissipative term $i\mu a_n$, com $n=1,2$, as shown in the system of equations 
(\ref{perturbada}). Initially we take $\mu=-0.05$, a weak dissipation. Note that in figures 4-6 the propagation shows attenuation and oscillations in the amplitude of the waves. These oscillations are explained by the transfer (coupling) of energy between the fundamental and second harmonic waves, due to the parameter $\beta$ in the equation system (\ref{maxwell}). Finally, it is shown in figure 6 that the soliton waves, in the presence of a linear dissipative term, undergo dispersion (pulse widening).  

\begin{figure}
\begin{center}
\caption{Propagation of modes $|a_1(\xi,s)|$ and $|a_2(\xi,s)|$, when subjected to linear dissipation $\mu=-0.05$.}
\subfigure[Mode $|a_1(\xi,s)|$]{\includegraphics[scale=0.37]{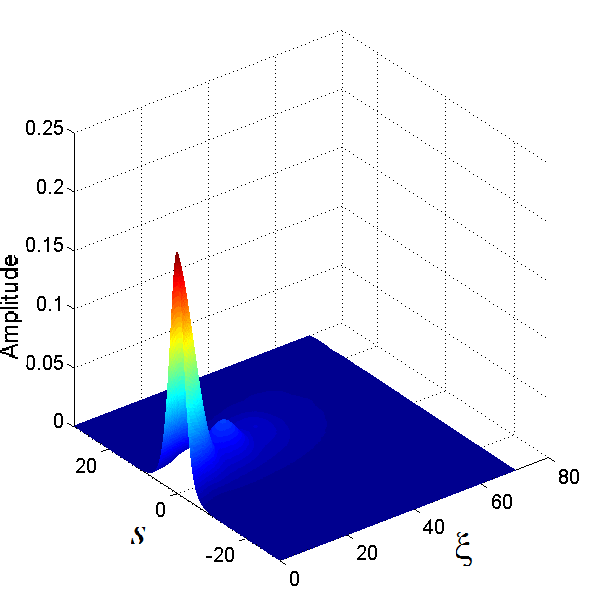}
  \label{}}\qquad \qquad
\subfigure[Mode $|a_2(\xi,s)|$]{\includegraphics[scale=0.37]{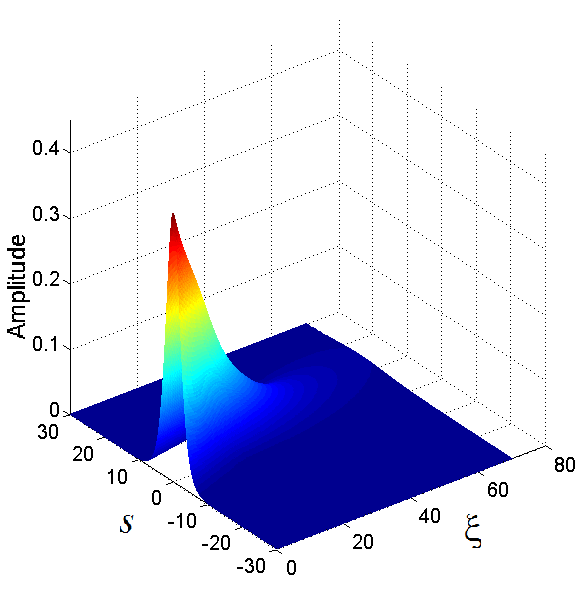}
   \label{}} 
\label{linear}  
\center{{\bf{Source:}} The Authors}
\end{center}
\end{figure}

\begin{figure}
\begin{center}
\caption{Profile of modes $|a_1(\xi,s)|$ and $|a_2(\xi,s)|$, when submitted to linear dissipation $\mu=-0.05$.}
\includegraphics[scale=0.25]{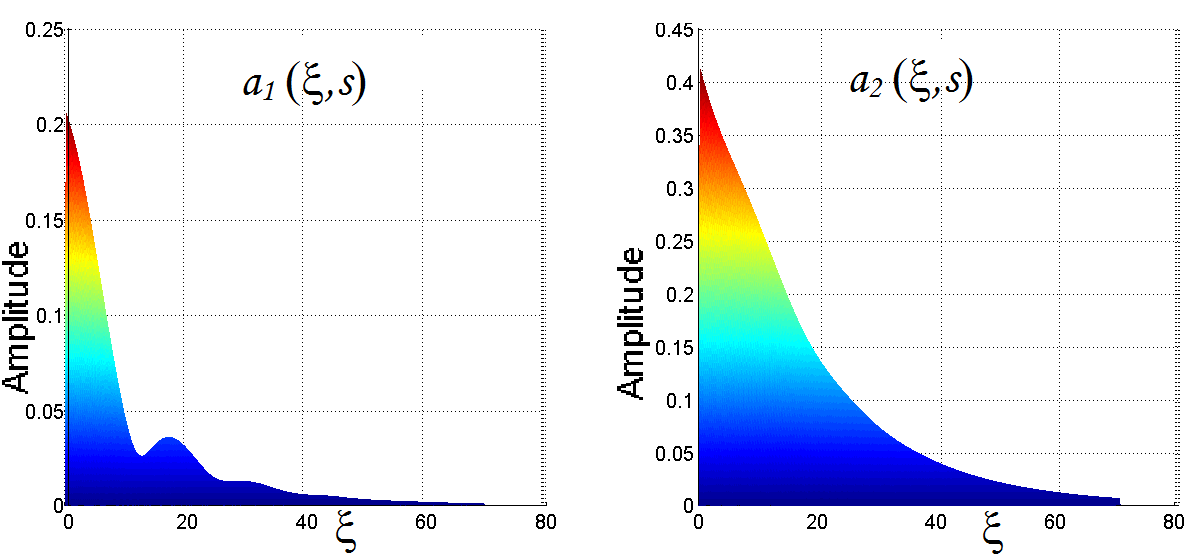}
\label{linear_perfil}  
\center{{\bf{Source:}} The Authors}
\end{center}
\end{figure} 

\begin{figure}
\begin{center}
\caption{Transverse section of modes $|a_1(\xi,s)|$ and $|a_2(\xi,s)|$ at $\xi=0$, $\xi=5$ and $\xi=18$, when subjected to linear dissipation $\mu=-0.05$.}
\includegraphics[scale=0.283]{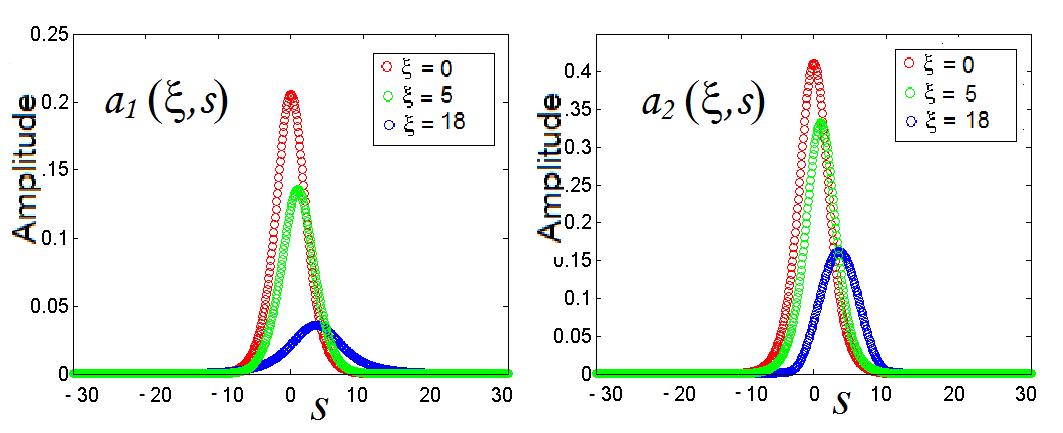}
\label{linear_cortes}  
\center{{\bf{Source:}} The Authors}
\end{center}
\end{figure}

In other simulations, when we take strong linear dissipations, for example $\mu=-1$, it is observed that the oscillations and dispersion effects in figures 4-5 disappear, since the intense attenuation factor masks this effect.

\begin{figure}
\begin{center}
\caption{Propagation of modes $|a_1(\xi,s)|$ and $|a_2(\xi,s)|$, when subjected to nonlinear dissipation $\kappa=-0.05$.}
\subfigure[Mode $|a_1(\xi,s)|$]{\includegraphics[scale=0.37]{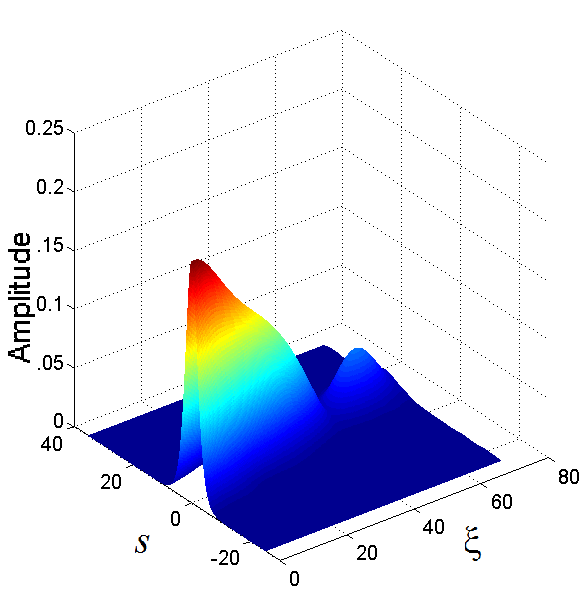}
  \label{}}\qquad \qquad
\subfigure[Mode $|a_2(\xi,s)|$]{\includegraphics[scale=0.37]{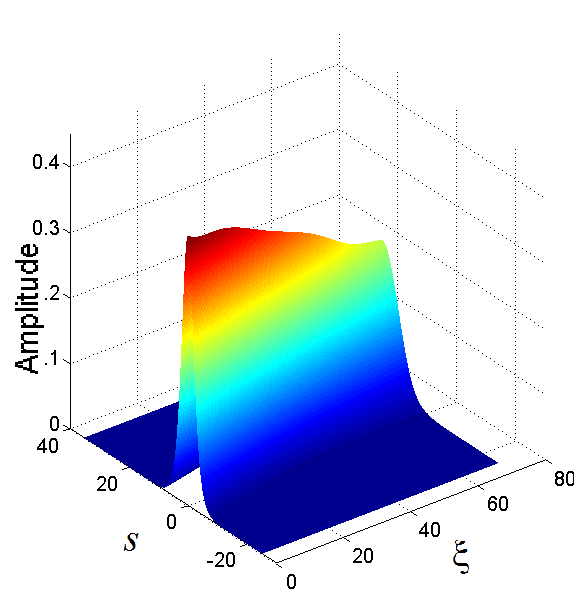}
   \label{}} 
\label{naolinear-1}  
\center{{\bf{Source:}} The Authors}
\end{center}
\end{figure}

\begin{figure}
\begin{center}
\caption{Profile of modes $|a_1(\xi,s)|$ and $|a_2(\xi,s)|$, when submitted to nonlinear dissipation $\kappa=-0.05$.}
\includegraphics[scale=0.25]{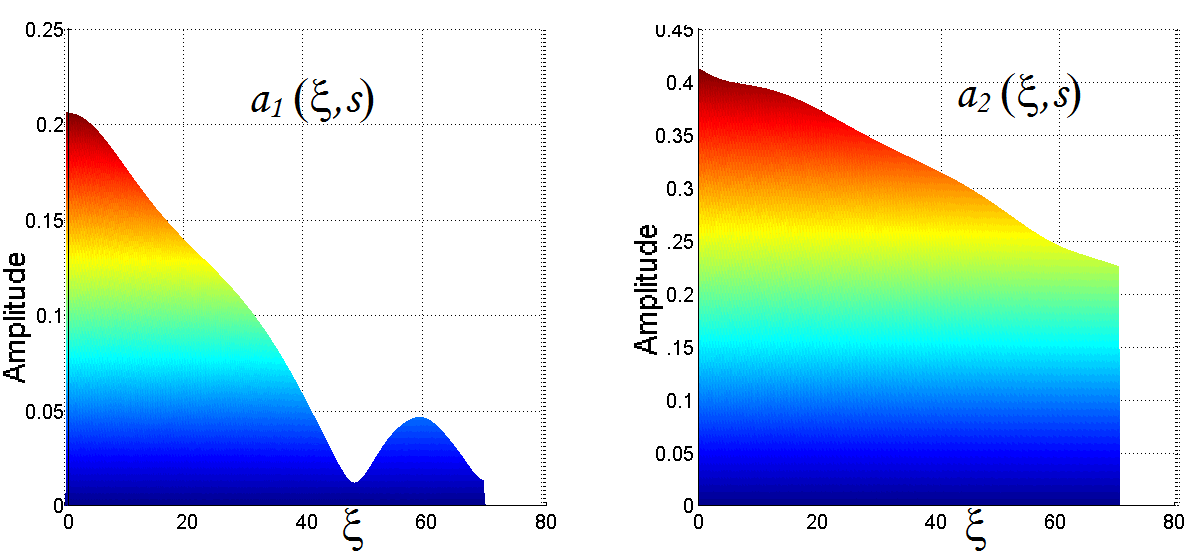}
\label{naolinear_perfil-1}  
\center{{\bf{Source:}} The Authors}
\end{center}
\end{figure}

\begin{figure}
\begin{center}
\caption{Transverse section of modes $|a_1(\xi,s)|$ and $|a_2(\xi,s)|$ at $\xi=0$, $\xi=30$ and $\xi=60$, when subjected to nonlinear dissipation $\kappa=-0.05$.}
\includegraphics[scale=0.283]{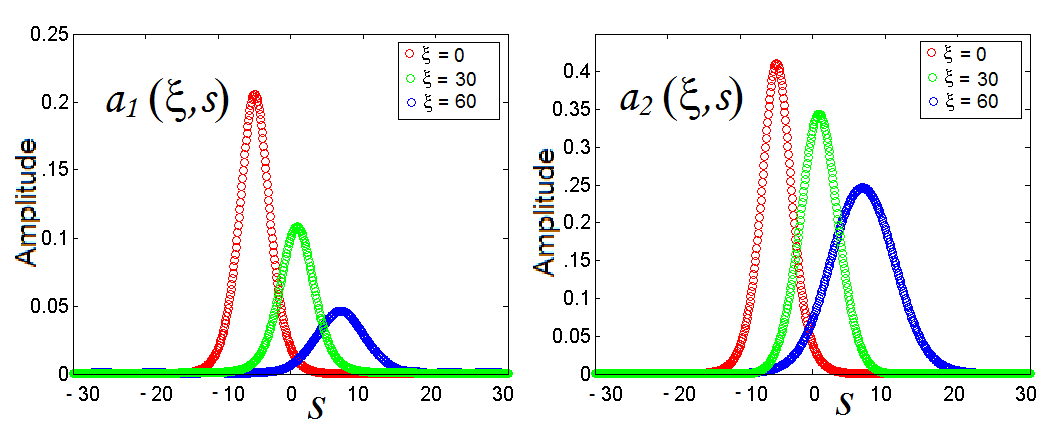}
\label{naolinear_cortes-1}  
\center{{\bf{Source:}} The Authors}
\end{center}
\end{figure}

In the next simulations we will study the nonlinear dissipation term $i\kappa |a_n|^{2} a_n$, with $n=1,2$. In figures \ref{naolinear-1} - \ref{naolinear_cortes-1} we take $\kappa=-0,05$. It is verified that decay rate with non-linearities is less intense than the decay rate due to linear dissipation ($\mu \neq 0$). The figure \ref{naolinear_perfil-1} shows oscillations in the amplitudes of $a_1(\xi,s)$ and $a_2(\xi,s)$. Note that the nonlinear dissipation term also generates dispersion in the pulse, figure \ref{naolinear_cortes-1}.

In other simulations, unlike in the case of linear dissipation,  when we take strong nonlinear dissipations, for example $\kappa=-1$, it is observed that the oscillations and dispersion effects in figures 8-9 increase. 

\subsection*{Numerical simulations for solitons with dissipation and gain}

In this subsection it is considered an optical fiber with linear losses of 1\%, that is, $\mu=-0,01$ in equation (\ref{perturbada}). It is also contemplated that in the link there is optical regeneration of the soliton signal through optical amplifiers, for example, erbium doped fiber amplifiers (EDFA), which allows amplifying the optical signal without the need for conversion of the optical-electric-optical signal (AGRA\-WAL, 2019).

In this context, we simulate the nonlinear properties that an optical fiber must have, so that the gain, given by $\kappa$, allows the soliton wave to triple its propagation distance, in relation to the situation in which there is only dissipation $\mu=-0,01$. We consider that the detection threshold at the receiver, due to noise, is 10\% of the initial amplitude.

\begin{figure}
\begin{center}
\caption{Propagation of modes $|a_1(\xi,s)|$ and $|a_2(\xi,s)|$, when subjected to linear dissipation $\mu=-0.01$.}
\subfigure[Mode $|a_1(\xi,s)|$]{\includegraphics[scale=0.37]{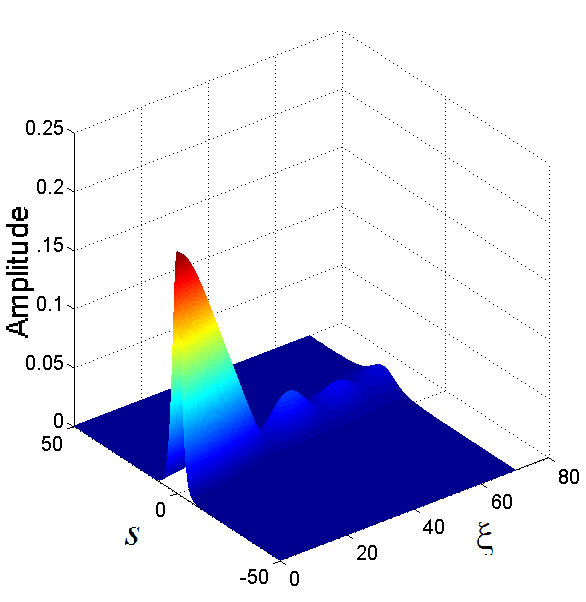}
  \label{}}\qquad \qquad
\subfigure[Mode $|a_2(\xi,s)|$]{\includegraphics[scale=0.37]{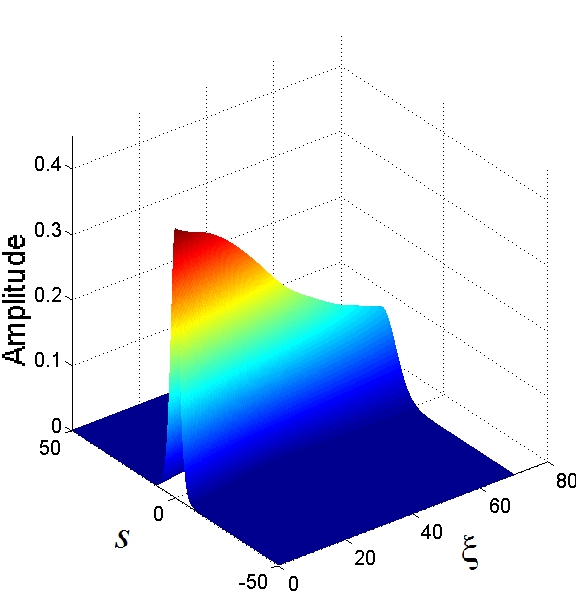}
   \label{}} 
\label{kappa0}  
\center{{\bf{Source:}} The Authors}
\end{center}
\end{figure}

\begin{figure}
\begin{center}
\caption{Profile of modes $|a_1(\xi,s)|$ and $|a_2(\xi,s)|$, when submitted to linear dissipation $\mu=-0.01$.}
\includegraphics[scale=0.25]{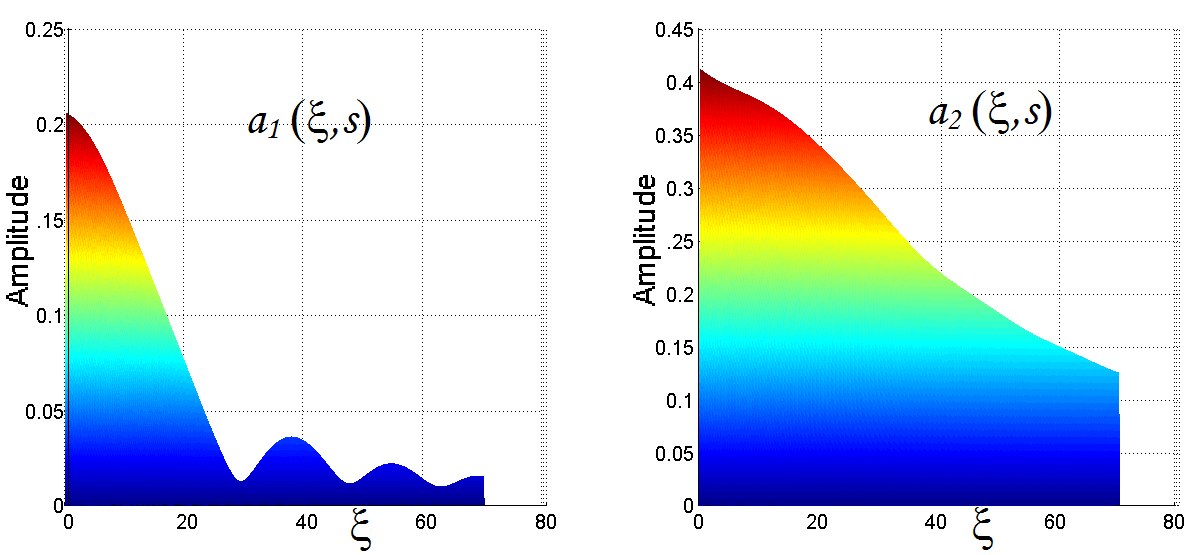}
\label{kappa0_perfil}  
\center{{\bf{Source:}} The Authors}
\end{center}
\end{figure}

\begin{figure}
\begin{center}
\caption{Transverse section of modes $|a_1(\xi,s)|$ and $|a_2(\xi,s)|$ at $\xi=0$, $\xi=40$ and $\xi=55$, when subjected to linear dissipation $\mu=-0.01$.}
\includegraphics[scale=0.283]{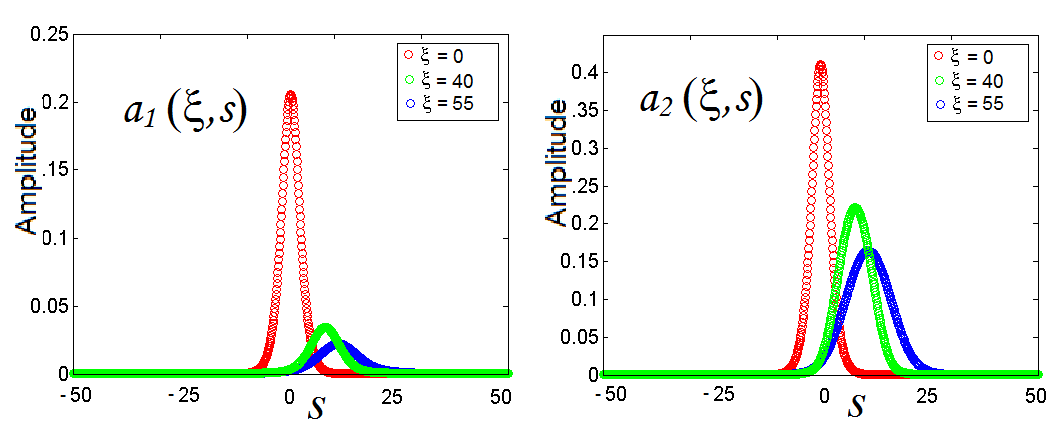}
\label{kappa0_cortes}  
\center{{\bf{Source:}} The Authors}
\vskip 0.5cm
\end{center}
\end{figure}

In the first simulation, represented in figures \ref{kappa0} - \ref{kappa0_cortes}, we consider the soliton wave only with linear dissipation $\mu=-0,01$ (with $\kappa=0$). Consistently with the previous results, we observed the attenuation, oscillation and dispersion of the fundamental and second harmonic modes.

Note that in the figure \ref{kappa0_perfil}, around $\xi=28$, the fundamental wave amplitude is less than 10\% of the initial amplitude. In this configuration the signal is not read by the detector, causing a transmission failure. In this context, we intend to simulate values for gain $\kappa$ in order to increase the propagated distance of the fundamental and second harmonic waves before this fault occurs. 

\begin{figure}
\begin{center}
\caption{Profile of modes $|a_1(\xi,s)|$ and $|a_2(\xi,s)|$, when submitted to linear dissipation $\mu=-0.01$ and nonlinear gain $\kappa=0,01$.}
\includegraphics[scale=0.25]{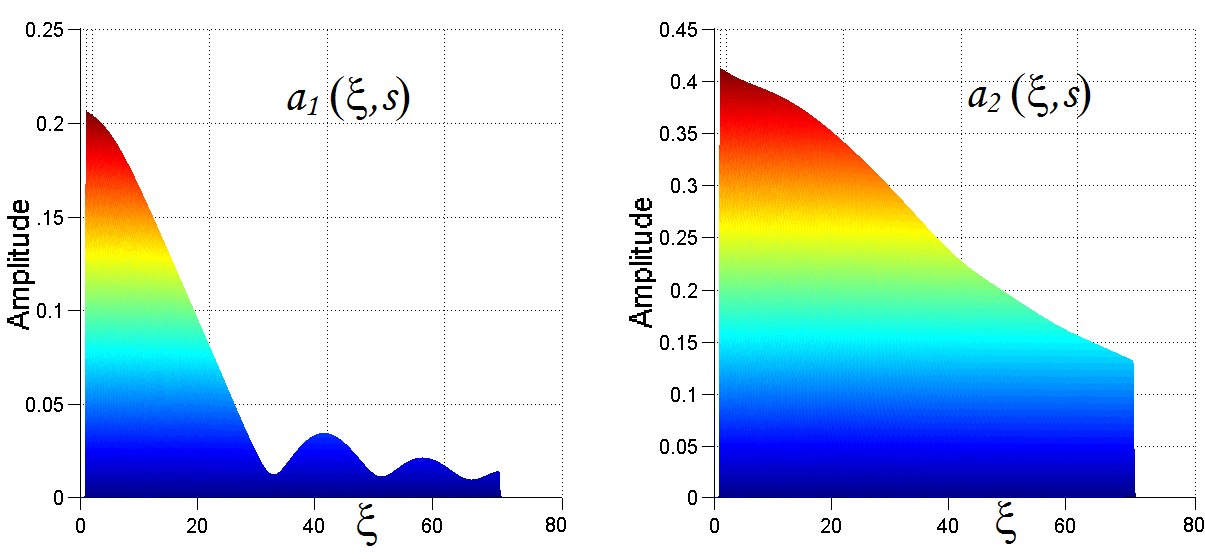}
\label{kappa1_perfil}  
\center{{\bf{Source:}} The Authors}
\vskip 0.5cm
\end{center}
\end{figure}

Below we consider the following gains: 1\%, 3\%, 6\%, and 9\%. Considering in the optical fiber a gain of 1\% ($\kappa=0,01$ with $\mu=-0,01$), it is observed in the figure \ref{kappa1_perfil} a small increase of the propagated distance of the fundamental wave. Now the first minimum occurs at approximately $\xi=33$.

\begin{figure}
\begin{center}
\caption{Profile of modes $|a_1(\xi,s)|$ and $|a_2(\xi,s)|$, when submitted to linear dissipation $\mu=-0.01$ and nonlinear gain $\kappa=0,03$.}
\includegraphics[scale=0.25]{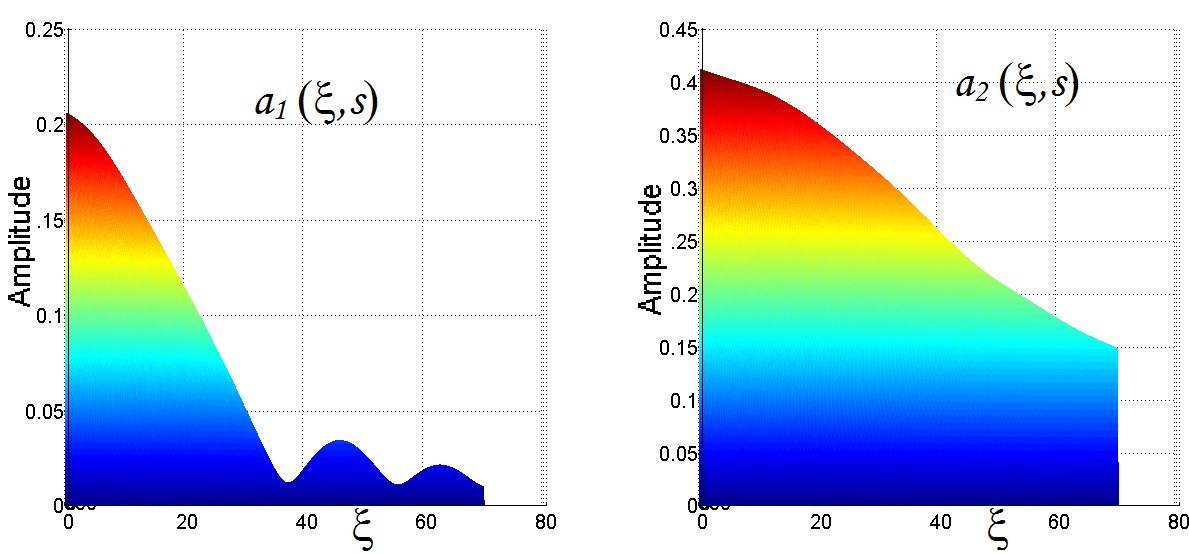}
\label{kappa3_perfil}  
\center{{\bf{Source:}} The Authors}
\end{center}
\end{figure}

\begin{figure}
\begin{center}
\caption{Profile of modes $|a_1(\xi,s)|$ and $|a_2(\xi,s)|$, when submitted to linear dissipation $\mu=-0.01$ and nonlinear gain $\kappa=0,06$.}
\includegraphics[scale=0.25]{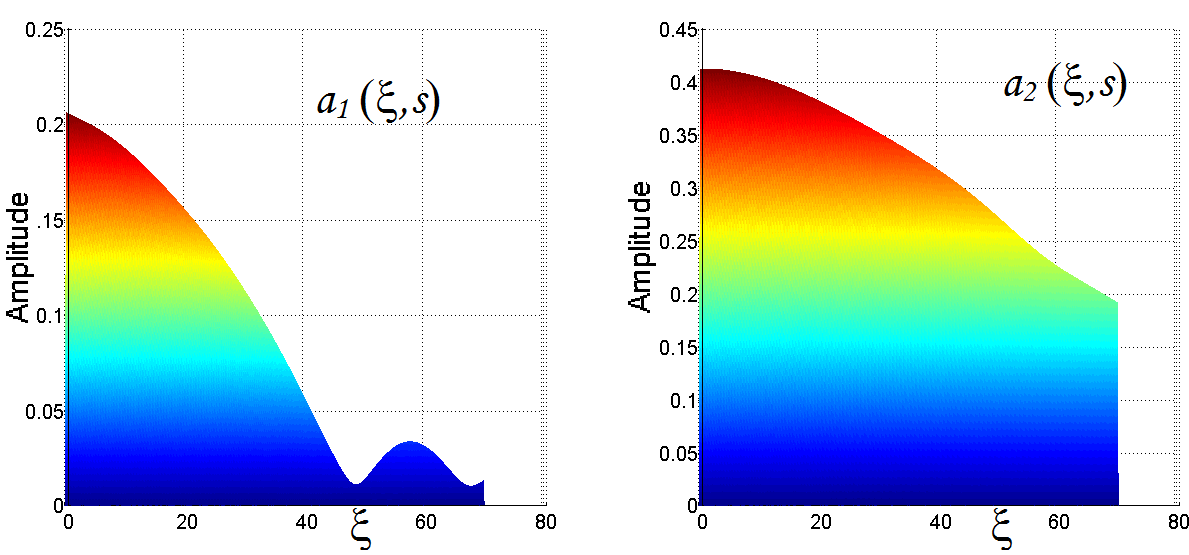}
\label{kappa6_perfil}  
\center{{\bf{Source:}} The Authors}
\end{center}
\end{figure}

The figures \ref{kappa3_perfil} and \ref{kappa6_perfil} show the fundamental and second harmonic mode propagation profiles for gains $\kappa=0,03$ and $\kappa=0,06$ (with $\mu=-0,01$). Note in the figures \ref{kappa3_perfil} and \ref{kappa6_perfil} a significant increase in the fundamental wave propagation distance before the first minimum occurs.  Note that the first minimum amplitude are located at $\xi=37$ e $\xi=48$, respectively. 

\begin{figure}
\begin{center}
\caption{Profile of modes $|a_1(\xi,s)|$ and $|a_2(\xi,s)|$, when submitted to linear dissipation $\mu=-0.01$ and nonlinear gain $\kappa=0,09$.}
\includegraphics[scale=0.25]{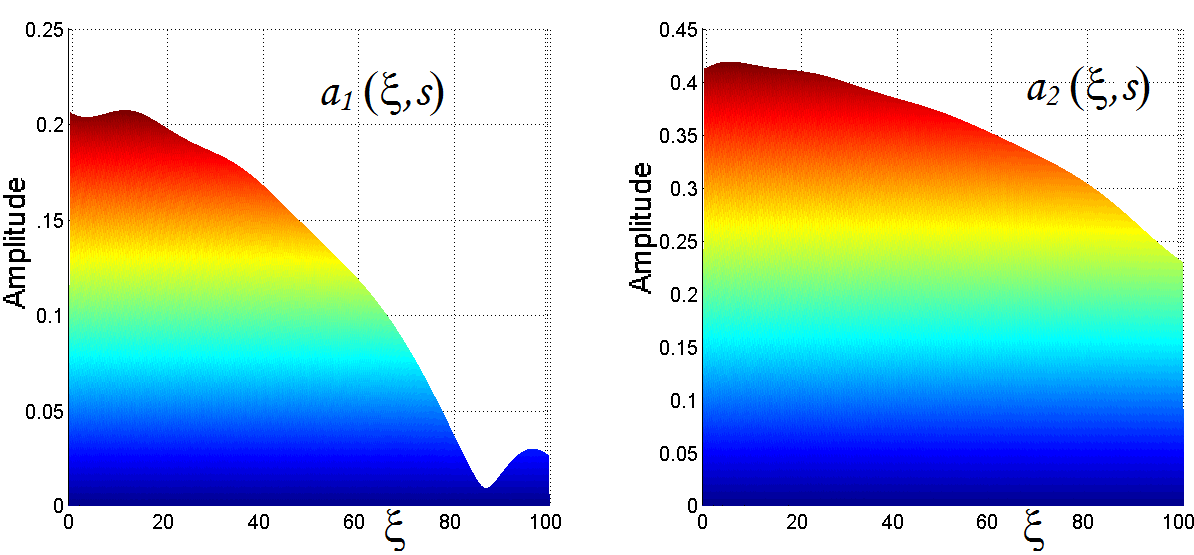}
\label{kappa9_perfil}  
\center{{\bf{Source:}} The Authors}
\end{center}
\end{figure}

Finally, the figure \ref{kappa9_perfil} simulates the situation corresponding to the gain $\kappa=0,09$. In this case, the first minimum amplitude of the fundamental wave occurs at approximately $\xi=88$. We conclude that with a gain of 9\%, it is possible to triple the propagation distance of the fundamental soliton wave, when subjected to linear dissipation of 1\%, before the first minimum occurs.

Note that in the situations analyzed in the figures \ref{kappa0_perfil}, \ref{kappa1_perfil}, \ref{kappa3_perfil}, \ref{kappa6_perfil}, and \ref{kappa9_perfil}, substantial attenuation in the fundamental wave amplitude occurs, while the attenuation of the second harmonic amplitude is less intense. In this context, we point out that, from an experimental point of view, it is more interesting to use the second harmonic wave as the information carrier wave, since all the dissipation and gain analysis made in this section for the fundamental wave is also valid for the second harmonic wave.

\section*{Conclusion}

In this work we numerically simulate the propagation of soliton waves in non-ideal optical fibers, subject to dissipation and gain.
 
Initially, we present the modeling of a nonlinear complex differential equation system (\ref{maxwell}) that describe the propagation of solitons waves in 
$\chi^{(2)}$ ideal fibers. The analytical solutions of the soliton type (2-3) were also discussed.

Then, through mathematical modeling, we add dissipation and gain terms to equations (1). The new equations no longer have soliton-type solutions. To simulate the propagation of these perturbed soliton waves, also called quasi-soliton waves, we adapt\-ed the numerical procedure developed in (CIRILO; NATTI; ROMEIRO; NATTI; OLIVEIRA, 2010), which employs the finite difference method associated with the relaxation Gauss-Seidel method, to solve the CPDE system (4). 

Right away, the numerical study of CPDE system showed that quasi-soliton waves suffer attenuation, oscillation and dispersion effects. Specifically, we were interested in the study of nonlinear optical signal amplification, the cubic term of the equation (4), when $\kappa>0$. We show that by this gain is possible to partially compensate the attenuation of the optical signal, described by the linear term of equation (4), when $\mu<0$.

We conclude that with a gain of 9 \%, it is possible to triple the fundamental soliton wave's progression distance, when subjected to linear dissipation of 1 \%, before the first minimum of its amplitude occurs. 

Another very significant result of this work was that, from an experimental point of view, it is more interesting to use the second harmonic wave as the carrier information wave, since it presents the dissipation and oscillation effects with less intensity than the fundamental wave, see figures \ref{kappa0_perfil}, \ref{kappa1_perfil}, \ref{kappa3_perfil}, \ref{kappa6_perfil}, and \ref{kappa9_perfil}.

As a suggestion for future work, we can improve the modeling of the gain and loss mechanisms in the propagation of soliton pulse. What are the intensities and patterns of disturbances caused in the optical pulse due to different gain and loss  mechanisms, for example, effects described by cubic terms (LATAS; FERREIRA, 2007) or derivative terms (KOHL; BISWAS; MILOVIC; ZERRAD, 2008)? This study would allow to numerically verify the effect on the signal caused by each mechanism, separately. 

\section*{Acknowledgments}

The author C. Dall'Agnol thanks the Londrina State University and CAPES for the scholarship granted from March/2012 to February/2014.

\section*{References}

\noindent AGRAWAL, G. P. {\it Nonlinear Fiber Optics}. 6th ed., San Diego: Academic Press, 2019.

\noindent ALGETY SITE. In: https://www.crunchbase.com/

\noindent organization/algety

\noindent ARTIGAS, D.; TORNER, L.; AKLMEDIEV, N. N. Dynamics of quadratic soliton excitation. {\it Optics Communications Journal}, v. 162, p. 347-356, 1999. 

\noindent ASHRAF, R.; AHMAD, M. O.; YOUNIS, M.; ALI, K.; RIZVI, S. T. R. Dipole and Gausson soliton for ultrashort laser pulse with high order dispersion. {\it Superlattices and Microstructures}, v. 109, p.504-510, 2017.

\noindent CHEMNITZ, M.; GEGHARDT, M.; GAIDA, C.; STUTZKI, F.; KOBELKE, J.; LIMBERT, J.; TUNNERMANN, A.; SCHMIDT, M. A. Hybrid soliton dynamics in liquid-core fibers. {\it Nature Communications}, v. 8, p. 42-52, 2017.

\noindent CIRILO, E. R.; NATTI, P. L.; ROMEIRO, N. M. L.; NATTI, E. R. T. Determination of the optimal relaxation parameter in a numerical procedure of solitons propagation. {\it Revista Ciências Exatas e Naturais}, v. 10, p. 77-94, 2008. 

\noindent CIRILO, E. R.; NATTI, P. L.; ROMEIRO, N. M. L.; NATTI, E. R. T.; OLIVEIRA, C. F. Soliton in ideal optical fibers – a numerical development. {\it Semina: Exact and Technological Sciences}, v. 31, p. 57-68, 2010.

\noindent CIRILO, E. R.; BARBA, A. N. D.; NATTI, P. L.; ROMEIRO, N. M. L. A numerical model based on the curvilinear coordinate system for the MAC method simplified. {\it Semina: Exact and Technological Sciences}, v. 39, p. 87-98, 2018.

\noindent CIRILO, E. R.; PETROVSKI, S. V.; ROMEIRO, N. M. L.; NATTI, P. L. Investigation into the critical domain problem for the Reaction-Telegraph equation using advanced numerical algorithms. {\it International Journal of Applied and Computational Mathematics}, v. 5, Article ID 54 (25pp), 2019.

\noindent EFTEKHAR, M. A.; EZNAVEH, Z. S.; AVILES, H. E. L.; BENIS, S.; LOPEZ, J. E. A.; KOLESIK, M.; WISE, F.; CORREA, R. A.; CHRISTODOULIDES, D. N.  Accelerated nonlinear interactions in graded-index multimode fibers. {\it Nature Communications}, v.10, p. 1638-1647, 2019.

\noindent FUKUCHI, K.; KASAMATSU, T.; MORIE, M.; OHHIRA, R.; ITO, T; SEKIYA, K; OGASAHARA, D.; ONO, T. 10.92-Tb/s (273 x 40-Gb/s) triple-band ultra-dense WDM optical-repeatered transmission experiment," in {\it Optical Fiber Communication Conference and International Conference on Quantum Information}, OSA Technical Digest Series, paper PD24, 2001. In: https://www.osapubli

\noindent shing.org/abstract.cfm?uri=OFC-2001-PD24

\noindent GALLÉAS, W.; YMAI, L.H.; NATTI, P.L.; NATTI, E. R. T. Solitons wave in dieletric optical fibers (in Portuguese). {\it Revista Brasileira de Ensino de Física}, v. 25, p. 294-304, 2003. 

\noindent ISMAIL, M. S. Numerical solution of coupled nonlinear Schrödinger equation by Galerkin method. {\it Mathematics and Computers in Simulation}, v. 78, p. 532-547, 2008.

\noindent ISMAIL, M. S.; ASHI, H. A. A compact finite difference schemes for solving the coupled nonlinear Schrodinger-Boussinesq equations. 
{\it Applied Mathematics}, v. 7, p. 605-615, 2016.

\noindent KARCZEWSKA, A.; ROZMEJ, P.; SZCZECINSKI, M.; BOGONIEWICZ, B. A. finite element me\-thod for extended KdV equations. {\it International Journal of Applied Mathematics and Computer Science}, v. 26, p. 555-567, 2016

\noindent KOHL R.; BISWAS, A.; MILOVIC, D.; ZERRAD, E. Optical soliton perturbation in a non-Kerr law media. {\it Optics \& Laser Technology}, v. 40, p. 647-662, 2008.

\noindent KUMAR, D. R.; RAO, B. P. Soliton interaction in birefringent optical fibers: Effects of nonlinear gain devices. {\it Optik}, v. 123, p. 117-124, 2012.

\noindent LATAS, S. C. V.; FERREIRA, M. F. S. Stable soliton propagation with self-frequency shift. {\it Mathematics and Computers in Simulation}, v. 74, p. 379-387, 2007.

\noindent LUO, J.; SUN, B.; JI, J.; TAN, E. L.; ZHANG, Y.; YU, X. High-efficiency femtosecond Raman soliton generation with a tunable wavelength beyond 2 $\mu$m. {\it Optics Letters}, v. 42, p. 1568-1571, 2017.

\noindent MENYUK, C. R.; SCHIEK, R.; TORNER L. Solitary waves due to $\chi^{(2)}$:$\chi^{(2)}$ cascading. {\it Journal of Optics of the Society American B – Optical Physics}, v. 11, p. 2434-2443, 1994.

\noindent OLIVEIRA, C. F.; NATTI, P. L.; CIRILO, E. R.; ROMEIRO, N. M. L.; NATTI, E. R. T. Numerical stability of solitons waves through splices in quadratic optical media. {\it Acta Scientiarum. Technology}, to appear in 2020.

\noindent PALMIERI, L.; SCHENATO, L. Distributed optical fiber sensing based on Rayleigh scattering. {\it The  Open Optics Journal}, v. 7, p. 104-127, 2013.

\noindent PARDO, S. R.; NATTI, P. L.; ROMEIRO, N. M. L.; CIRILO, E. R. A transport modeling of the carbon-nitrogen cycle at Igapó I Lake – Londrina, Paraná State, Brazil. {\it Acta Scientiarum. Technology}, v. 34, p. 217-226, 2012.

\noindent QUEIROZ, D. A.; NATTI, P. L.; ROMEIRO, N. M. L.; NATTI, E. R. T. A numerical development of the dynamical equations of solitons in optical fibers (in Portuguese). {\it Semina: Exact and Technological Sciences}, v. 27, p. 121-128, 2006.

\noindent ROMEIRO, N. M. L.; CASTRO, R. G. S.; CIRILO, E. R.; NATTI, P. L. Local calibration of coliforms parameters of water quality problem at Igapó I lake, Londrina, Paraná, Brazil. {\it Ecological Modelling}, v. 222, p. 1888-1896, 2011.

\noindent ROMEIRO, N. M. L.; MANGILI, F. B.; COSTANZI, R. N.; CIRILO, E. R.; NATTI, P. L. Numerical simulation of BOD5 dynamics in Igapó I lake, Londrina, Paraná, Brazil: Experimental measurement and mathematical modeling. {\it Semina: Exact and Technological Sciences}, v. 38, p. 50-58, 2017. 

\noindent SAITA, T. M.; NATTI, P. L.; CIRILO, E. R.; RO\-MEIRO, N. M. L.; CANDEZANO, M. A. C.; ACU\-NA, R. B.; MORENO, L. C. G. Simulação numérica da dinâmica de coliformes fecais no lago Luruaco, Colômbia. {\it Tendências em Matemática Aplicada e Computacional}, v. 18, p. 435-447, 2017.

\noindent SMITH, N. J.; KNOX, F.M.; DORAN, N. J.; BLOW, K. J.; BENNION, I. Enhanced power solitons in optical fibres with periodic dispersion management. {\it Electronics Letters}, v. 32, p. 54-55, 1996.

\noindent SMITH, G. D. Numerical Solution of Partial Differential Equations: Finite Difference Methods. New York: Oxford University Press 2004.

\noindent TAYLOR, J. R. {\it Optical Solitons Theory and Experiment}. Cambridge: Cambridge University Press, 1992.

\noindent TRIKI, H.; BISWAS, A.; MILOVIC, D.; BELIC, M. Chirped femtosecond pulses in the higher-order nonlinear Schrodinger equation with non-Kerr nonlinear terms and cubic-quintic-septic nonlinearities. {\it Optics Communications}, v. 366, p. 362-369, 2016.

\noindent WANG, W. C.; ZHOU, B.; XU, S. H.; YANG, Z. M.; ZHANG, Q. Y. Recent advances in soft optical glass fiber lasers. {\it Progress in Material Sciences}, v. 101, p. 90-171, 2019.

\noindent WEN, B.; YANGBAO, D.; SHI, X.; FU, X. (2018). Evolution of finite-energy Airy pulse interaction with high-power soliton pulse in optical fiber with higher-order effects. {\it Optik}, v. 152, p. 61-68, 2018.

\noindent YAMAI, L. H.; GALLÉAS, W.; NA\-TTI, P. L.; NA\-TTI, E. R. T. Stability of solitons in $\chi^{2}$-type dielectric optical fibers (in Portuguese). {\it Revista Ciências Exatas e Naturais}, v. 6, p. 9-29, 2004.

\noindent YUSHKO, O. V.; REDYUK, A. A.; FEDORUK, M. P.; TURITSYN, S. K. Coherent soliton communication lines. {\it Journal of Experimental and Theoretical Physics}, v. 119, p. 787-794, 2014.

\noindent ZAJNULINA, M.; BOHN, M.; BODENMULLER, D.; BLOW, K.; BOGGIO, J. M. C.; RIEZNIK, A. A.; ROTH, M. M. Characteristics and stability of soliton crystals in optical fibres for the purpose of optical frequency comb generation. {\it Optics Communications}, v. 393, p. 95-102, 2017.

\end{document}